\documentclass[twocolumn,showpacs,preprintnumbers,amsmath,amssymb,prb]{revtex4}
\usepackage{graphicx}% Include figure files
\usepackage{dcolumn}% Align table columns on decimal point
\usepackage{bm}% bold math

\begin{document}
\title{Ab-initio calculation method for charged slab systems \\
 using field-induced gaussian sheet}

\author{Seiji Kajita$^{1,2}$}
 \email{fine_controller@phys8.s.chiba-u.ac.jp}
\author{Takashi Nakayama$^1$}
\author{Maki Kawai$^{2,3}$}
\affiliation{
$^1$ Department of Physics, Faculty of Science, Chiba University, Yayoi
Inage, Chiba 263-0022, Japan
}
\affiliation{
 $^2$ RIKEN (The Institute of Physics and Chemistry Research), 2-1
Hirosawa, Wako, Saitama 351-0198, Japan
}
\affiliation{
$^{3}$Department of Advanced Materials Science, Graduate School of
Frontier Sciences, University of Tokyo, Kashiwa, Chiba 277-8561
}
\date{\today}

\begin{abstract}
A new repeated-slab calculation method is developed to simulate the
electronic structures of charged surfaces by arranging 
density-variable charged sheets in vacuum regions to realize a 
constant potential on the charged sheets and maintain the 
charge neutrality condition. 
The charged sheets are fabricated so as to screen an electric field from charged
slabs; consequently, they act like a counter electrode composed of flat 
perfect conductors, modeling
a tip of a scanning tunneling microscope or a reference electrode in
an electrochemical cell.
This method has the advantages of ease 
of implementation into existing repeated-slab programs and low
computational costs. 
The availability of the method is demonstrated by applying it
to a charged H$_2$ molecule and charged Al(111) metal and Si(111)
semiconductor surfaces.
\end{abstract}

\pacs{Valid PACS appear here}% PACS, the Physics and Astronomy
                             % Classification Scheme.
\maketitle
\section{Introduction} %% No sections necessary for express letters, letters and short notes
A surface becomes an interesting stage for material transformations 
such as chemical reactions when the surface is placed under an 
external electric field and has positive/negative charges. 
In the scanning tunneling microscope (STM), for example, 
the applied sample bias produces an electric field between the surface 
and the STM tip.\cite{WHo}
Such electric field often induces a variety of surface reconstructions,
\cite{Takagi_Komori} desorption,\cite{Onishi_Iwasawa} 
and the diffusion/rotation of
adsorbates.\cite{Carpinelli_Swartzentruber,
Swartzentruber_Jonsson}
On the other hand, in electrochemistry, the manipulation of electric 
fields is a key factor in controlling chemical 
reactions.\cite{Kolb, Lipkowski}

Ab-initio calculation using a repeated slab has greatly contributed 
to the exploration of the geometric and electronic structures of a
number of surfaces. 
Such calculation employs a periodic boundary condition (PBC) 
perpendicular to the slab; thus, it receives the benefit of numerous
computational techniques developed in the calculations of periodic bulk
systems. 
However, most of these calculations are concerned with neutral surfaces,
because various physical quantities such as energy and potential 
diverge to infinity unless charge neutrality is maintained within a unit 
cell in an electronic structure calculation based on the PBC scheme.\cite{Makov} 

To treat charge-neutral surfaces in an external electric field, 
Neugebauer and Scheffler\cite{Neugebauer_Scheffler} introduced 
artificial dipole sheets in vacuum regions of a repeated unit cell.
Arranging the positive and negative charged sheets 
on the front and back sides of the slab, respectively, one can easily 
impose a constant electric field on the slab system. 
The calculation method for a charged slab was performed
by Fu and Ho. \cite{Fu_Ho}
In their method, artificially charged sheets, which have 
the same amount but an opposite sign of charge to that of the slab, 
are placed on both sides of the slab. 
These artificially charged sheets terminate the electric field 
originating from the charged slab. 
All these computational recipes dexterously realize 
the charge neutrality and compensate the electric field in a unit cell,
and they have been utilized in several studies for the elucidation of field
effects on surfaces.\cite{Mattson_Feibelman, Feibelman, Lozovoi, HeChe}

However, since these methods use sheets with fixed charge distributions,
typically of plane Gaussian form, the electric field from a charged slab
 is in general not perpendicular 
to the sheet and the potential shows a modulation along the sheet. 
This feature becomes marked and produces some inconvenience 
when the surface of the slab is bumpy and the excess-charge distribution on 
the slab is far from uniform. 
For example, unless the thickness of the vacuum region is sufficiently large, 
the potential modulation on the sheet often produces unnecessary 
electrostatic interactions between repeated slabs, which makes 
it complicated to analyze the energetics on the slab. 
Furthermore, in the case of a STM experiment or an electrochemical cell, 
since a STM tip and a reference electrode
are normally made of conducting materials, they have a constant 
potential but a nonuniform charge distribution, 
rather than a fixed uniform charge distribution (see Fig. 1(a)). 
To simulate the situations in which such a counter conducting material
 approaches a surface, it is desirable to adopt another calculational 
recipe using constant-potential charged sheets. 

Recently, Otani and Sugino\cite{Otani_Sugino} have developed a general method 
of simulating charged surfaces surrounded by screening mediums. 
They ingeniously combined a repeated-slab calculation with a
real-space potential calculation to respond to any non-PBCs
 by releasing the charge-neutrality constraint of a unit cell. 
However, their method needs the regeneration of a potential in real space 
and requires additional computational costs. 

The purpose of this work is to demonstrate a new computational recipe for 
simulating a charged surface located near a counter metallic 
electrode maintaining the charge-neutrality constraint by only 
inserting a constant-potential charged sheet within the
repeated-slab scheme.
Our approach provides the advantages of ease of implementation 
into existing ab-initio calculation programs based on PBC and
 low additional computational costs for the charged slab calculation.
The rest of this paper is organized as follows: 
in \S \ref{sec2}, we explain how to construct a constant-potential
charged sheet and then present formulas to realize it 
using an existing ab-initio calculation program. 
In \S \ref{sec3}, we demonstrate how the present method works, 
by showing results for a charged H$_2$ molecule, charged Al(111) and 
Si(111) surfaces near an electrode. 
An attention with respect to the width of charged sheets is also presented. 
We give a brief summary of this paper in \S \ref{sec4}.

\section{\label{sec2}Formalism}

In this section, we consider how to construct a charged sheet 
that has a constant potential on the sheet.
When one connects a conducting material and a surface with an 
electric battery, 
counter charge is induced on the conducting material 
by an electric field from the charged surface, 
and this realizes a constant potential on a conducting material,
as shown in Fig. 1(a). \cite{electrodynamics}
To construct the induced counter charge, we first obtain the electric field
from a charged slab.

%%%%%%%%%%%%%%%%%%%%%%%%%%%%%%%%%%%%%%%
\begin{figure*}[htbp]
 \begin{center}
 \includegraphics[width=.8\linewidth]{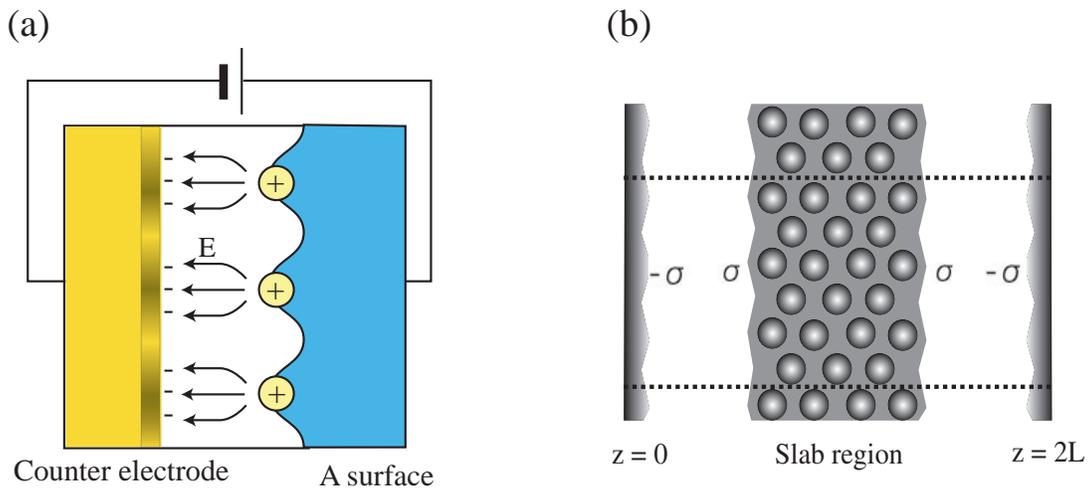}
  \caption{\label{pic:setup}
(a) Schematic picture of charged surface system considered in this work. 
The surface and the counter electrode are connected with 
a cathode and an anode of an electric battery, respectively. 
Arrows denote an electronic field, $\textbf{E}$, in the vacuum region. 
(b) A slab setup employed in the present calculation to 
simulate the charged surface shown in (a). 
The center of the slab is located at $z=L$ and each surface 
of the slab has a charge $+\sigma$. 
Two counter charged sheets are set at both boundaries 
and each electrode has the counter charge $-\sigma$.
}
 \end{center}
\end{figure*}
%%%%%%%%%%%%%%%%%%%%%%%%%%%%%%%%%%%%%%%

We prepare an open space that is periodic in both $x$ and $y$ directions but
nonperiodic in $z$ direction,
where the $x$, $y$ and $z$  directions correspond to the surface lateral 
and normal directions, respectively.
The range of the $z$ direction is limited to 
$0 \le z \le 2L$, and the center of the charged slab is located at $z=L$, 
as shown in Fig. \ref{pic:setup}(b). 
We assume that the slab has reflection symmetry at $z=L$, 
for simplicity.
The charged slab produces an electric field outside the slab, 
\textit{ i.e.}, in the vacuum regions. 
The $z$ component of such an electric field at the boundaries, $z^*=0$ 
or $z^*=2L$, is obtained as 
\begin{eqnarray}
E_z(z^*,\textbf{g}) 
= - \int_0^{2L}  \partial_z G(z=z^*, z', \textbf{g}) \rho_{slb}(z',
\textbf{g}) 
dz',  
\label{eq:rho-G0}
\end{eqnarray}
where we performed a Fourier transformation along the $x-y$ direction 
and used the reciprocal component, $\textbf{g}$. 
$\rho_{slb}(z, \textbf{g})$ is the electronic density 
of electrons and ions in the slab region.
G(z, z',\textbf{g}) is the propagating Green's function obtained by
solving the Fourier-transformed Poisson equation, 
$ (\partial_z^2 -g^2) G(z, z',\textbf{g}) = -4\pi \delta(z-z'),$ 
where $g=|\textbf{g}|$. 
If we set the electronic potential to zero at the cell boundaries, 
 $G(z=z^*, z', \textbf{g})=0$, 
this Green's function can be expressed as 
\begin{eqnarray}
\lefteqn{G(z, z',\textbf{g})=\frac{2\pi}{g} e^{-g|z-z'|} +
 \frac{2\pi}{g} } 
 \nonumber \\
&&\times \frac{ e^{-2Lg} \cosh(g(z-z')) -\cosh(g(z+z'-2L))
  }{\sinh(2Lg)}. \nonumber \\
\label{eq:deltafunG0}
\end{eqnarray}

Then, we arrange the charged sheets at the boundaries, 
$z^*=0$ and $z^*=2L$. 
If the charge density on the sheet is determined so as to 
compensate $E_z(z^*,\textbf{g})$,
a constant potential is realized at the cell boundaries
and the charge neutrality is naturally satisfied in $0 \le z \le 2L$ 
on the basis of Gauss's theorem. 
Therefore, we employ $E_z(z^*,\textbf{g})$ to determine the charge densities on the 
sheets and adopt 
the half-Gaussian forms for the charge distribution along the $z$ direction.

Since the above-mentioned cell, which is composed of the charged slab and counter 
charged sheets at the cell boundaries, is neutral and produces no electric field 
outside the cell, we stack such cells periodically along the $z$ direction. 
In this case, the repeated counter charge density, 
$\rho_{F}(z, \textbf{g})$, is given by 
\begin{eqnarray}
 \rho_{F}(z, \textbf{g}) &=& -\frac{E_z(z^*=2L, \textbf{g})}{4\pi A} \nonumber \\
&\times& \sum_{n=-\infty}^{\infty} 
e^{-(\frac{z-2L(n+1)}{a})^2} \theta(2L(n+1)-z) \nonumber \\
&+& \frac{E_z(z^*=0, \textbf{g})}{4\pi A} \nonumber \\
&\times& \sum_{n=-\infty}^{\infty} 
e^{-(\frac{z-2Ln}{a})^2}  \theta(z-2nL), 
\label{eq:realGSperiodic}
\end{eqnarray}
where $\theta$ and $a$ are the step function and the width of a Gaussian 
charge, respectively. 
$A$ is a normalization constant of a half-Gaussian distribution, 
$A\equiv \int_{0}^{\infty} e^{-(\frac{z}{a})^2}=\frac{\sqrt{\pi}}{2}a$. 
Inserting eqs. (\ref{eq:rho-G0}) and (\ref{eq:deltafunG0}) into 
eq. (\ref{eq:realGSperiodic}) and Fourier-transforming the 
result along the $z$ direction, we obtain for $g \ne 0$ 
\begin{eqnarray}
\rho_{F}(k, \textbf{g}) =
-e^{-(\frac{ka}{2})^2} \frac{g}{4\pi L}\tanh(Lg) 
\sum_{k'} \phi_{slb}(k', \textbf{g})  \nonumber \\
- \frac{1}{4\pi L} sgn(k)\ erf(|k|a/2) 
 \sum_{k'} \phi_{slb}(k', \textbf{g}) k',
\label{eq:lastrho21}
\end{eqnarray}
and for $g = 0$ 
\begin{eqnarray}
\lefteqn{\rho_{F}(k,\textbf{g}=\textbf{0}) = -e^{-(\frac{ka}{2})^2} n_0 }
\nonumber \\
&&- \frac{sgn(k)\ erf(\frac{|k|a}{2}) }{4\pi L} 
 \sum_{k'\neq 0} \phi_{slb}(k', \textbf{g}=0) k', 
\label{eq:lastrho22}
\end{eqnarray}
where $k$ is the reciprocal component along the $z$ direction and 
$\phi_{slb}(k, \textbf{g})$ is the Coulomb potential due to the slab charge, 
$4 \pi \frac{\rho_{slb}(k, \textbf{g})}{k^2+\textbf{g}^2}$. 
$n_0$ represents the excess electron number in a slab given by
$n_0 = \int_{unit} \rho_{slb}(\textbf{r}) d\textbf{r}$. 
Positive and negative $n_0$'s correspond to an
excess and a deficiency of electrons in the slab, respectively. 
Hereafter, we call this type of counter charged sheet a
``field-induced Gaussian-charge (FIGC) sheet''.

In the repeated-slab calculation, the total charge density is 
obtained as the sum of the charge density of the FIGC sheet, 
$\rho_{F}(k, \textbf{g})$, and that of the slab, $\rho_{slb}(k, \textbf{g})$. 
Such a sum enables the electronic-structure calculation of 
charged surfaces located nearest to a constant-potential electrode 
by a standard \textit{ab-initio} method based on the PBC. 
By only adding $\rho_{F}(k, \textbf{g})$, we can obtain physical
quantities such as the electron distribution, the electronic potential, 
forces on ions, and the total energy of the system 
composed of the charged surface with the contour electrode. \cite{energycom}

\section{\label{sec3}Results and discussion}

In this section, we demonstrate how the present scheme works. 
We consider a charged H$_2$ molecule, and charged Al(111) and Si(111) 
$1 \times 1$ surfaces. 
These systems are arranged at the center of a repeated unit cell and 
the atomic positions of all systems are optimized. 
We perform a standard \textit{ab-initio} calculation 
based on the density functional theory (DFT)\cite{Hohenberg_Kohn, Kohn_Sham}
using a plane-wave basis set with an energy less than 25 Ryd. for the
H$_2$ system and 16 Ryd. for the Al(111) and 
Si(111) systems. 
The present scheme is implemented into the computational code 
``{\small Tokyo ab-initio program package}'' (Tapp).\cite{TAPP}
The Perdew-Burke-Ernzerhof (PBE)\cite{PBE1} functional in a 
 generalized gradient approximation (GGA) is employed as an 
exchange-correlation potential for the H$_2$ system, 
while the Perdew-Wang\cite{PW92} functional in a 
local density approximation (LDA) is adopted for the Al(111) and 
Si(111) systems. 
The ionic cores are described by ultrasoft 
pseudopotentials\cite{Vanderbilt} in all cases. 

\begin{figure*}[htbp]
 \begin{center}
 \includegraphics[width=.8\linewidth]{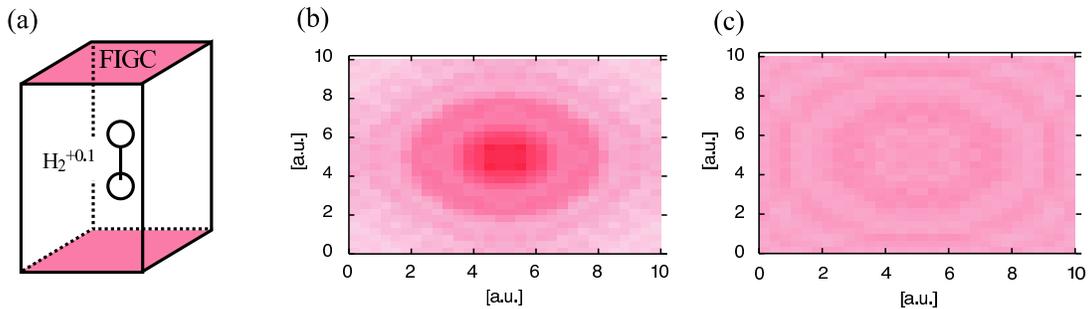}
  \caption{\label{pic:h2figs}
(a); Schematic view of charged H$_2^{+0.1}$ molecule arranged in 
 repeated unit cell. 
FIGC indicates the position of field-induced Gaussian-charge sheets, 
which are located at both the top and bottom of the cell. 
(b) and (c) are contour maps of charge distribution on the sheets 
when the sheets are located (b) $5.5$ and (c) $8.7$ a.u.
 from a nearest hydrogen atom. 
Denser color corresponds to higher density.
}
 \end{center}
\end{figure*}

Figure \ref{pic:h2figs}(a) shows the setup of a H$_2$ molecule 
in a unit cell. 
Unit cells having $10.0 \times 10.0 \times 12.5$ and 
$10.0 \times 10.0 \times 18.75$ a.u.$^3$ sizes are adopted. 
The H$_2$ molecule is surrounded 
by vacuum in all directions. 
The counter charged sheets are set at unit cell boundaries of the z direction. 
We ionize this molecule positively as H$_2^{+0.1}$ by 
extracting 0.1 electrons. 
Figures \ref{pic:h2figs}(b) and \ref{pic:h2figs}(c) show the 
counter maps of induced charge density calculated at the cell boundaries, 
when the charged sheets are placed $5.5$ and $8.7$ a.u. 
apart from the nearest hydrogen atom of the molecule, respectively. 
In Fig. \ref{pic:h2figs}(c), since the distance between 
the molecule and the counter charged sheet is large,
the electric field generated by the molecule uniformly touches 
the sheet and the charge distribution on the sheet becomes uniform. 
However, when the sheet approaches the molecule, 
reflecting the nonuniform charge distribution of H$_2^{+0.1}$, 
the induced charge shows a spacial modulation, as shown 
in Fig. \ref{pic:h2figs}(b).

%%%%%%%%%% H esgs appearance  %%%%%%%%%
%%%%%%%%%%%%%%%%%%% 

\begin{figure}[htbp]
 \begin{center}
 \includegraphics[width=.75\linewidth]{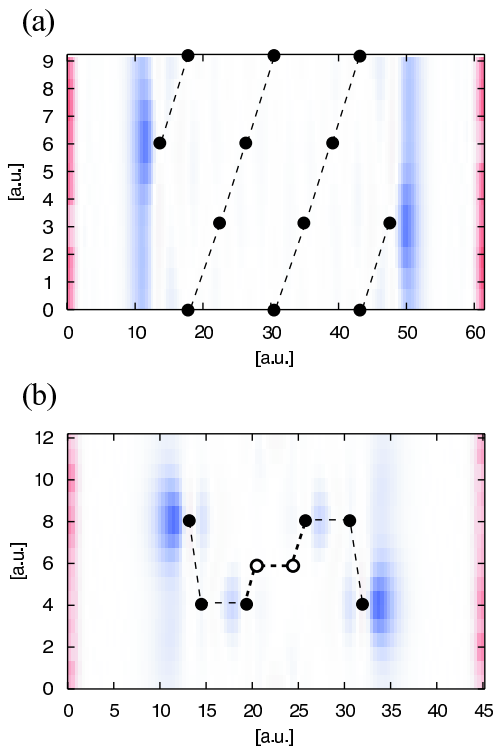}
  \caption{  \label{pic:cutexcess}
Cross-sectional [$10\bar{1}$] and [$1\bar{1}0$]
 views of excess-charge distributions for 
(a) Al$^{+0.1}$(111) and (b) Si$^{+0.1}$(111) surface slabs. 
The charge distributions of the counter charged sheets are also shown. 
Blue and red areas correspond to positive and negative charges, respectively. 
Solid circles indicate atoms on the display plane, 
while open circles indicate atoms out of the plane. 
A Gaussian-charge width of $a=1.0$ a.u. is adopted in both (a) and (b).
}
 \end{center}
\end{figure}

Next, we consider positively charged Al(111) and Si(111) surfaces. 
Figures \ref{pic:cutexcess}(a) and \ref{pic:cutexcess}(b) show 
the excess-charge distributions of these systems 
viewed from the [$10\bar{1}$] and [$1\bar{1}0$] directions, respectively. 
Both surfaces are located about 10 a.u. from 
the counter electrodes. 
The excess-charge distribution is calculated as the difference in 
the charge density between charged and neutral systems. 
In the case of the Al(111) metal surface, since the electrons are weakly bound 
to atoms in metals, the excess charge is uniformly distributed along 
the surface. 
As a result, the counter charge on the sheet becomes uniform an parallel 
to the surface. 
On the other hand, in the case of the Si(111) surface, 
because dangling bonds have the highest energy in the surface,
the inserted positive charge is mainly localized around dangling 
bonds of surface Si atoms.
Thus, the counter charge on the sheet focuses at positions just 
above and below the dangling bonds. 

%%%%%%%%%% sAl & sSi cut  %%%%%%%%%
%%%%%%%%%%%%%%%%%%%

Then, we consider how a constant zero potential is realized on 
the counter charged sheet. 
In the present repeated-slab scheme, 
there exists one artificial parameter:
the width of the Gaussian-charge sheet, denoted as $a$ in 
eq. (\ref{eq:realGSperiodic}). 
This width is introduced to decrease the number of 
the plane-wave basis set and, thus, the computational time. 
However, owing to a finite $a$, the zero-potential condition 
assumed in the present method is slightly released. 
This situation can be understood by observing the potentials
on the sheet. 
The potential at $z=0$ is, for example, obtained by the sum of 
the $k$-components of the total potential as 
$\phi(\textbf{g}, z=0) = \sum_{k=-\infty}^{\infty} 
\phi(\textbf{g}, k)/2L$. 
Using a simple calculation, we obtain 
\begin{eqnarray}
\lefteqn{
\phi(\textbf{g}, z=0)= 
\frac{1}{2L} \sum_{k'=-\infty}^{\infty} 
\phi_{slb}(\textbf{g},k')}
  \nonumber \\
&&\times \big\{ 
1-\frac{g}{L}\tanh(Lg)\sum_{k=-\infty}^{\infty}\frac{e^{-(\frac{ka}{2})^2}}{k^2+g^2}
\big\}
\label{eq:conditionsum1}
\end{eqnarray}
At $a=0$, it can be analytically derived that
the right hand of eq.(\ref{eq:conditionsum1}) equals
zero using the series formula
$\sum_{n=-\infty}^{\infty} \frac{1}{X^2+n^2} = \frac{\pi}{X \tanh(\pi
X)}$,
where $X$ is an arbitrary real number and $n$ is an integer.
Hence, the zero-potential condition is satisfied on the sheet. 

\begin{figure}[htbp]
 \begin{center}
 \includegraphics[width=.85\linewidth]{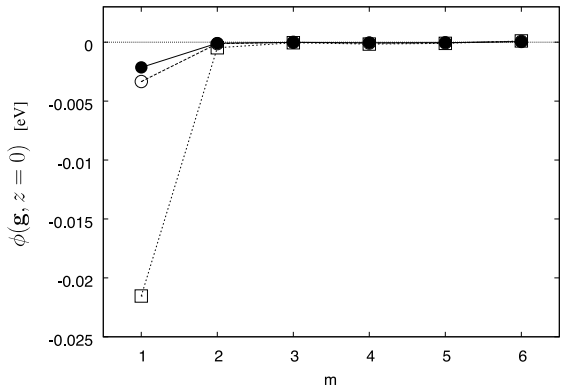}
  \caption{\label{pic:vboundH}
Fourier components of total potential on counter charged sheet 
at $z=0$, $\phi(\textbf{g}, z=0)$, for various in-plane wave vectors, 
$\textbf{g}$. 
A charged $H^{+0.1}_2$ molecule is located at the center of a unit cell, 
5.5 a.u. apart from the sheets. 
We adopt $\textbf{g}$ characterized by the an integer $m$, 
as $\textbf{g}={\rm g}_0 (m, m)$, where ${\rm g}_0$ is the length of the
in-plane fundamental reciprocal vector.
Solid and open circles indicate the potential
components obtained by the FIGC scheme using the widths $a=0.3$ and 
$a=0.5$ a.u., respectively, whereas open squares denote those 
obtained by the conventional fixed Gaussian-charge-sheet 
method using a width of $a=0.5$ a.u..
}
 \end{center}
\end{figure}

On the other hand, in the case of a finite $a$, 
we show in Fig. \ref{pic:vboundH} the calculated $\phi(\textbf{g}, z=0)$ 
for the charged H$_2^{+0.1}$ system shown in Fig. \ref{pic:h2figs}(a). 
Here, the charged sheets are arranged 5.5 a.u. apart from the hydrogen
molecule. 
The $\phi(\textbf{g}, z=0)$ values using the conventional 
fixed Gaussian-charge-sheet method are also shown for reference. 
This $\phi(\textbf{g}, z=0)$ represents the spacial modulation of 
potential on the counter charged sheets, 
corresponding to the wave vector \textbf{g}. 
The result using the fixed Gaussian-charge sheet
shows a potential modulation of about 20 meV, thus, 
not realizing the constant zero potential. 
On the other hand, in the case of the present scheme, 
it is seen that the potential modulation of each wave vector is 
suppressed and 
the potential modulation decreases with $a$.
Therefore, a zero potential can be realized 
as long as a small $a$ is employed.
%%%%%%%%%%%%%%%%%%%%%%%%%%%%%%%%%%%%%%%
%%%%%%%%%%%%%%%%%%%%%%%%%%%%%%%%%%%%%%%

\section{\label{sec4}Summary}

To calculate the electronic structures of charged surfaces 
located near a flat and metallic counter electrode, 
we developed a new method that uses neutralizing variable charged sheets
to realize a constant potential on the electrode. 
Since the present method is constructed on the basis of the PBC and 
a neutral unit cell, 
this method has the significant advantages that the implementation into 
existing repeated-slab programs is easy and 
the additional computational time is short.
We have demonstrated that the present scheme works well 
not only for metal surfaces but also for semiconductor and 
insulator surfaces, and surfaces with adsorbates. 
We hope that the present scheme will be implemented in various 
programs and used for the study of charged surfaces.

\begin{acknowledgments}
This work was supported by the Junior Research Associate (JRA) 
fellowship of RIKEN, 
the Ministry of Education, Culture, Sports, Science and 
Technology of Japan,
the RIKEN Research Program ``Nanoscale Science and Technology Research''
and the 21 COE program of Chiba University. 
We thank the RIKEN Super Computing Center (RSCC)
and Chiba University for the use of facilities. 
S.K. wishes to acknowledge Dr. O. Sugino and Dr. M. Otani for many
helpful discussions and comments.
\end{acknowledgments}
%%%%%%%%%%%%%%%%%%%%%%%%%%%%%%%%%%%%%%%

%
%\bibliography{main}% Produces the bibliography via BibTeX.
%
\end{document}